\documentclass[12pt]{iopart}
\usepackage{hyperref}
\usepackage{amsfonts}

\begin{document}

\title[Kolmogorov-Sinai and Bekenstein-Hawking entropies]{Kolmogorov-Sinai and Bekenstein-Hawking entropies}
\author{Kostyantyn Ropotenko}
\address{State Department of communications and
informatization, Ministry of transport and communications of
Ukraine, 22, Khreschatyk, 01001, Kyiv, Ukraine}

\begin{abstract}
It is shown that instability of stringy matter near the event
horizon of a black hole (the spreading effect) can be characterized
by the Lyapunov exponents. The Kolmogorov-Sinai entropy is the sum
of all the positive Lyapunov exponents and equals to the inverse
gravitational radius. Due to a replacement of the configuration
space of a string by its phase space at distance of order of the
string scale, the relation between the Kolmogorov-Sinai and
Bekenstein-Hawking entropies is established. The KS entropy of a
black hole measures the rate at which information about the state of
a string collapsing into the black hole is lost with time as it
spreads over the horizon.
\end{abstract}


\bigskip\bigskip
\maketitle

\section{Introduction}

It is well known that some general relativistic systems described by
the Einstein equations can exhibit chaotic behavior \cite{bar,hob}.
One of the most important quantities characterizing the chaotic
behavior of a dynamical system is the Kolmogorov-Sinai (KS) entropy,
which describes the rate of change of information about the phase
space trajectories as a system evolves (a more formal definition
will be given below). On the other hand, some general relativistic
systems possess thermal properties and can be characterized by the
Boltzmann, or the thermodynamical entropy. In particular, black
holes being the solutions of the Einstein equations are
characterized by the Bekenstein-Hawking (BH) entropy of just the
same kind. In this connection, an important question arises: is
there a relation between the KS and BH entropies?

In this paper I propose a possible relation between the KS and BH
entropies. In the following sections we will introduce the main
conceptions of chaotic dynamics and black hole thermodynamics,
demonstrate chaotic behavior of stringy matter near the event
horizon of a black hole, and  establish the relation between the KS
and BH entropies.

\section{The main conceptions of chaotic dynamics and black hole thermodynamics}

We begin with definitions. Suppose that phase space of a dynamical
system is finite, then the KS entropy $h_{KS}$ is the sum of all the
positive Lyapunov exponents of the system, where the Lyapunov
exponents $h_{i}$ characterize the rate of exponential separation of
the nearby system's trajectories in phase space as a result of a
local instability \cite{shu}

\begin{equation}
\label{tr} d(t)=d(0)\:e^{h_{i}t}.
\end{equation}
If this leads to an increase in the phase space volume occupied by
the system with time
\begin{equation}
\label{ph} \Delta{\Gamma(t)}=\Delta{\Gamma(0)}\:e^{h_{KS}t},
\end{equation}
the Boltzmann entropy increases \cite{zas}
\begin{equation}
\label{B} S(t)=h_{KS}t+\ln(\Delta{\Gamma(0)}).
\end{equation}
As is easily seen, the KS entropy $h_{KS}$ is not really an entropy
but an entropy per unit time, or entropy rate, $dS/dt$. Note that
the linear relation between the KS and thermodynamical entropies is
not a general case \cite{lat, fal}.

The BH entropy of a black hole, on the other hand,
\begin{equation}
\label{BH}S_{BH}=\frac{A}{4G}=\frac{\pi R_{g}^{2}}{l_{P}^{2}}\:,
\end{equation}
is obtained from the thermodynamical relation $dE=TdS$, where the
energy of the black hole is its mass $M$, the temperature is given
by $T=1/8\pi GM$, and the area of the event horizon $A$ is related
with the gravitational radius $R_{g}$, $R_{g}=2GM$, in the usual way
$A=4\pi R_{g}^{2}$. The BH entropy is defined in the reference frame
of an external distant observer at fixed static position above the
horizon (an external observer).

Our purpose is to find a kinematic effect caused by the black hole
geometry with respect to which a system evolves as in Eqns.
(\ref{tr}),(\ref{ph}) in the reference frame of an external
observer. For this purpose we repeat, for completeness, some
well-known facts from \cite{s} concerning the behavior of matter
near the horizon without proofs, thus making our exposition
self-contained.

\section{A classical particle near the event horizon of a black hole}

First consider a classical particle falling toward the horizon. The
main fact is that the proper time in the frame of the particle
$\tau$ and the Schwarzschild time of an external observer $t$ are
related through $\tau \sim e^{-t/2R_{g}}$  due to the redshift
factor. Therefore in order to observe the particle before it crosses
the horizon, we have to do it in a time which is exponentially small
as $t\rightarrow \infty$. In other words, an external observer sees
the particle as being slowed down with increasing powers of
resolution. Moreover in the frame of the observer, the momentum of
the particle increases like $\sim e^{t/2R_{g}}$, so that the
observer sees the particle as being flattened in the direction of
motion due to Lorentz longitudinal contraction.

The other important fact is that if the particle behaves as a
conventional classical object it will appear to have fixed
transverse size on the horizon. In so doing, the phase space volume
holds its shape and remains the same; all information is conserved:

\begin{equation}
\label{clas} \Delta{\Gamma_{clas}(t)}=\Delta{\Gamma_{clas}(0)}.
\end{equation}

\section{Chaotic behavior of a relativistic string collapsing into a black hole}

My proposal rests on stringy matter having unusual kinematic
properties near the event horizon of a black hole. According to
string theory, the most promising candidate for a fundamental theory
of matter, all particles are excitations of a one-dimensional object
- a string. String theory is characterized by two fundamental
parameters: the string scale, $l_{s}$, and the string coupling
constant $g$; if $l_{P}$ is the Planck length then $l_{P}=g\:
l_{s}$. An important fact is that strings behave very differently
from ordinary particles. The crucial difference is that the size and
shape of a string are sensitive to the time resolution. Susskind has
shown \cite{su} that the mean squared radius of a string, $\langle
R_{s}\rangle ^{2}$ depends on the time resolution, $\tau _{r}$ as
$\langle R_{s}\rangle ^{2}\sim \ln (1/ \tau _{r})$ for $\tau _{r}\ll
1$.

Consider a string falling toward a black hole. As mentioned above,
an external observer has a time resolution that decreases like
$e^{-t/2R_{g}}$. This means that the string approaching the event
horizon spreads in the transfers directions in the reference frame
of the observer like $t/2R_{g}$ (there is also a longitudinal
spreading but it is rapid to balance the Lorentz longitudinal
contraction). Thus the string, in contrast to the classical
particle, will not appear to have fixed transverse size on the
horizon. As we have seen, the growth of the string is linear. But as
noted by Susskind himself \cite{s}, this result was obtained in the
framework of free string theory. It doesn't take into account such a
nonperturbative phenomenon as string interactions; there are
indications \cite{s}-\cite{mez} that a true growth must be
exponential
\begin{equation}
\label{rad}\langle R_{s}\rangle ^{2}\sim e^{t/R_{g}}.
\end{equation}
This also means that close trajectories of bits of the string
diverge exponentially
\begin{equation}
\label{tra} d_{bit}(t)=d_{bit}(0)\:e^{t/2R_{g}}.
\end{equation}

In addition, Susskind \cite{suss} and Mezhlumian, Peet and
Thorlacius \cite{mez} have found that string configuration becomes
chaotic and very complicated like a fractal during the spreading
process. They have shown that as the correlation length of a string
decreases exponentially with time the number of bits of the string
increases exponentially
\begin{equation}
\label{bit}N_{bit} \sim e^{t/2R_{g}}.
\end{equation}
They have interpreted this as a branching diffusion process, where
every bit diffuses independently of others over the whole horizon
and bifurcates into two bits and so on. According to the authors the
diffusion process should provide necessary thermalization as the
string spreads over the horizon.

But this picture also permits another interpretation. First the
diffusion is a distinctive random process. But in our case there are
no real random forces. The behavior of a string near the horizon is
very well described by the Hamilton dynamics. If there are exact
equations of motion no true randomness is possible.  Second the
string is a fundamental object. It is not a dissipative system. In
the spreading process no points of a string are lost and also no
points are gained: the number of bits of a string is conserved.
Kinetics of the diffusion process is based on the random phase
approximation, which implies rapid decay of correlations in the
system. Chaotic dynamics of a string, on the other hand, gives the
finite mixing time (see below Eqn. (\ref{time})), which just means a
finite decay time of the correlations. So chaotic dynamics ensures
the important condition of randomness that is crucial for deriving
of diffusion kinetics. Therefore we can give the following
interpretation of Eqn. (\ref{bit}). Initially bits of a string
occupy one cell in phase space of a string. In the course of time,
all bits will move to different phase-space points, mapping the cell
at time $t=0$ to another cell at time $t$. Hence we can interpret
Eqn. (\ref{bit}) as an increase in the number of occupied cells
\begin{equation}
\label{cel}N_{cell}(t) = N_{cell}(0)\: e^{t/2R_{g}}.
\end{equation}
But this number is  proportional to the distance between the
trajectories of bits that all initially occupy one cell (\ref{tra}),
as required.

The spreading process begins to occur when the string reaches the
horizon at distance of order of the string scale $l_{s}$ from the
horizon in a thin layer $\sim l_{s}$. But in string theory at such
scales the mirror symmetry should takes place \cite{pol, mar}. In
general it relates the complex and K\"{a}hler structures of some
manifolds. In the simplest case for closed strings it exchanges the
winding number around some circle with the corresponding momentum
quantum number (T-duality) or, roughly speaking, coordinates with
momenta. At the scales $\gg l_{s}$ we can always single out the
configuration space and the phase space is its cotangent bundle. At
the scales $\sim l_{s}$ this is not the case: at such scales there
is a replacement of the configuration space of a string by its phase
space \cite{mar}. A similar phenomenon in quantum mechanics - a
particle in magnetic field \cite{lan}: on the distances of order of
the magnetic length $l_{mag} \sim \sqrt{\hbar c/eH}$ a replacement
of the configuration plane transversal to the direction of the
magnetic field by the phase plane takes place so that the number of
states is $A/l^{2}_{mag}$, where $A$ is the area of the transversal
plane.

\section{The KS entropy of a black hole and its relation with the BH entropy}

Hence instead Eqn. (\ref{clas}) we obtain
\begin{equation}
\label{str} \Delta
{\Gamma_{s}(t)}=\Delta{\Gamma_{s}(0)}\:e^{t/R_{g}}.
\end{equation}
Then, taking into account Eqns. (\ref{rad})-(\ref{str}), we conclude
that the spreading effect realizes  a two-dimensional flow (or map)
on the horizon by means of the positive Lyapunov exponents, $h_{i}=
1/2R_{g}; i=1,2$. Thus string matter collapsing into a black hole
exhibit chaotic behavior which can be characterized by the KS
entropy
\begin{equation}
\label{KSB} h_{KS}=\frac{1}{R_{g}}\:.
\end{equation}
Note that $h_{KS}$ is infinite in purely random systems \cite{shu}.

Finally we can obtain the relationship between the KS and BH
entropies. Since $\Delta{\Gamma_{s}(t)}=4\pi R_{g}^{2}$ and in the
strong coupling regime ($g\sim 1$, $l_{s}\sim l_{P}$)
$\Delta{\Gamma_{s}(0)}= l_{P}^{2}$ (or the same $\langle
d_{bit}(t)\rangle ^{2}=4\pi R_{g}^{2}$ and $\langle
d_{bit}(0)\rangle ^{2}= l_{P}^{2})$, we have
\begin{equation}
\label{KSBH} h_{KS}= \frac{d (\ln S_{BH})}{d t}\:,
\end{equation}
where $S_{BH}$ is identified with the string entropy and expressed
in terms of the characteristic time of the black hole $R_{g}/c$.
Susskind has shown \cite{sus} that all black hole states are in
one-to-one correspondence with single string states. This agrees
with our identification.

The KS entropy $h_{KS}$  of a dynamical system measures the rate at
which information about the state of the system is lost with time.
We can determine the average time over which the state of a string
(or any body made of strings) can be predicted. Since the entire
accessible phase space of the string is bounded by the horizon area,
the trajectories of bits (\ref{tra}) mix together. This occurs when
\begin{equation}
\label{time} t_{mix}\sim R_{g}.
\end{equation}
At this time the string spreads over the entire horizon and can no
longer expand due to the nonperturbative effects \cite{s,su,suss}.
The result is crucial for the relaxation of the string to
statistical equilibrium: to reach a statistical equilibrium in a
finite time we should have the finite time of mixing (\ref{time}).
After the time $t_{mix}$ all information contained in the string
will be lost and we will able only to make statistical predictions.
This time is comparable to the characteristic time of a black hole
$R_{g}$ but is smaller than the black hole lifetime $\sim
R_{g}^{3}$. Thus the KS entropy $h_{KS}$  of a black hole measures
the rate at which information about the state of a string (or any
body made of strings) collapsing into the black hole is lost with
time as it spreads over the horizon.

We have demonstrated a relation between the KS and BH entropies for
a string spreading over the event horizon of a black hole. It is
widely believed, however, that the spreading effect is not a
peculiar feature of a special (still hypothetical) kind of matter.
In the framework of the so-called infrared/ultraviolet connection
\cite{s} it is a general property of all matter at energies above
the Planck scale. If we want to study progressively smaller and
smaller objects we must, according to conventional physics, to use
higher and higher energy probes. But once gravity is involved that
rule is changed radically. Since at energies above the Planck scale
black holes are created, it follows \cite{s} that as we raise the
energy we probe larger and larger distances. In other words very
high frequency is related to large size scale, $\Delta x\ \Delta
\tau \sim l_{P}^{2}$. Then, taking into account the redshift factor,
we can obtain the exponential growth of the transverse size of
matter similar to Eqn. (\ref{rad}), as required.

In conclusion, let us turn to the form of the relation between the
KS and BH entropies (\ref{KSBH}). It is interesting, to what extent
it is special and can one obtain a similar relation from the general
reasoning? For this purpose let us express the Boltzmann entropy not
in terms of phase volume (\ref{ph}) but in terms of a distribution
function $f(x,t)$
\begin{equation}
\label{ent} S=-\int f \ln f\:d\Gamma.
\end{equation}
Now suppose that near equilibrium $f(x,t)$ can be presented in the
form \cite{bar,chir}
\begin{equation}
\label{dis} f(x,t)=f_{eq} + (f_{0}(x)-f_{eq})e^{-h(t)t}.
\end{equation}
By differentiating Eqn. (\ref{ent}) with respect to $t$ and using
Eqn. (\ref{dis}) we obtain
\begin{equation}
\label{entr} \frac{\partial S}{\partial t}= \int
h\:(f(x,t)-f_{eq})\ln f \:d\Gamma.
\end{equation}
For short times $f\ll f_{eq}$ and Eqn. (\ref{entr}) reduces to
\begin{equation}
\label{entro} \frac{\partial S}{\partial t}= S_{eq}\int
h(x)\:d\Gamma.
\end{equation}
Then, since $h_{KS}=\int h(x)\:d\Gamma $ we have
\begin{equation}
\label{KS} h_{KS}=\frac{1}{S_{eq}}\frac{\partial S}{\partial
t}\approx \left(\frac{\partial \ln S}{\partial t}\right)_{S\simeq
S_{eq}},
\end{equation}
as required.

\section{The KS entropy of other spaces with the event horizon}

Of course, besides the black holes there are other general
relativistic systems, which possess thermal properties, and de
Sitter space is the most known of them. As is well known, it is a
thermodynamical system with the Gibbons-Hawking (GH) entropy given
by
\begin{equation}
\label{GH}S_{GH}=\frac{A}{4 l_{P}^{2}}=\frac{\pi}{H^{2} l_{P}^{2}
}\,,
\end{equation}
where $H$ is the Hubble constant, and the area of the event horizon
$A$ is related with the radius of de Sitter space $R_{dS}$,
$R_{dS}=H^{-1}$, in the usual way $A=4\pi R_{dS}^{2}$. We can repeat
our experiment with a string by throwing it toward the event horizon
of de Sitter space. Obviously, the result will be the same: the
string spreads over the horizon. Thus, repeating the previous
arguments, we can obtain the KS entropy of de Sitter space
\begin{equation}
\label{KSG} h_{KS}= H,
\end{equation}
and the relationship between the KS and GH entropies
\begin{equation}
\label{KSGH} h_{KS}=\frac{d (\ln S_{GH})}{d t}\:.
\end{equation}

\section{Conclusions}

In this paper we have shown that stringy matter near the event
horizon of a black hole with the gravitational radius $R_{g}$
exhibits instability (the spreading effect), which can be
characterized by the Lyapunov exponents. The Kolmogorov-Sinai
entropy is the sum of all the positive Lyapunov exponents, $h_{KS} =
1/R_{g}$. Due to a replacement of the configuration space of a
string by its phase space at distance of order of the string scale,
the relation between the Kolmogorov-Sinai and Bekenstein-Hawking
entropies is established, $h_{KS}=\partial(\ln S_{BH})/\partial t$,
where the black hole entropy is identified with the string entropy
and expressed in terms of the characteristic time of the black hole
$R_{g}/c$. The KS entropy of a black hole measures the rate at which
information about a string (or any body made of strings) collapsing
into a black hole is lost as the string (the body) spreads over the
horizon. Since the mixing time is finite $\sim R_{g}$, the system
reach a statistical equilibrium in a finite time.

\section*{References}
\bibliographystyle{iopart-num}

\end{document}